\documentclass[12pt]{iopart}

\usepackage{graphicx} 

\begin{document}

\title[About possible contribution of intrinsic charm component]{About possible contribution of
intrinsic charm component to inclusive spectra of charmed mesons}

\author{E.~V.~Bugaev and P.~A.~Klimai}

\address{ Institute for Nuclear Research, Russian Academy of Sciences,
60th October Anniversary Prospect 7a, 117312 Moscow, Russia }

\ead{pklimai@gmail.com}

\begin{abstract}
We calculate differential energy spectra ($x_F$-distributions) of charmed particles
produced in proton-nucleus collisions, assuming the existence of intrinsic heavy
quark components in the proton wave function. For the calculation, the recently proposed
factorization scheme is used, based on the Color Glass Condensate theory and specially
suited for predictions of a production of particles with large rapidities. It is
argued that the intrinsic charm component can, if it exists, dominate in a 
sum of two components, intrinsic + extrinsic, of the inclusive spectrum of charmed 
particles produced in proton-nucleus collisions
at high energies, in the region of medium $x_F$, $0.15 < x_F < 0.7$, and can give noticeable 
contribution to atmospheric fluxes of prompt muons and neutrinos.
\end{abstract}



\submitto{\JPG}
\maketitle

\section{Introduction}

According to the usual assumption, heavy quark partons in the nucleon are radiatively generated,
i.e., they appear only as a result of QCD evolution which starts from a null value at a scale of
approximately the corresponding quark mass. There are two reasons, at least, for this assumption:
{\it i)} if quark mass is large enough, the partonic probability distribution function (PDF) is perturbatively
calculable and {\it ii)} up to now, exact experimental constraints on these PDFs are absent.

The mass of charm quark ($\sim 1.3$ GeV) lies in between the soft and hard scales, so, the standard
method of PDF's calculation is not well justified. Besides, from a theoretical point of view the
existence of an "intrinsic charm" (IC) component of the nucleon is not forbidden.
Many nonperturbative models (the light-cone models \cite{Brodsky:1980pb, Brodsky:1981se},
the meson cloud picture \cite{Navarra:1995rq, Paiva:1996dd},
the MIT bag model \cite{Donoghue:1977qp}, $SU(4)$ quark model \cite{Song:2001ri}, etc.) predict
such a component at an energy scale comparable to mass of charm quark $m_c$.

In the present paper we use the predictions of the specific light-cone model developed in \cite{Brodsky:1980pb}
("BHPS model"). In \cite{Brodsky:1980pb} the PDF of charm quarks from the $5$-quark component $uudc\bar c$
in the proton is calculated. The $uudc\bar c$-Fock state arises in QCD not only from gluon
splitting which is included in DGLAP evolution, but also from diagrams in which the heavy quark
pair is multi-connected to the valence constituents. The latter components are called
intrinsic charm Fock components. It was originally suggested in \cite{Brodsky:1980pb} that
there is $\sim 1\%$ probability of IC Fock states in the nucleon. The materialization of the
IC Fock state (after its interaction with the target) leads to the production of open-charm
states such as scalar mesons $D^+(c\bar d)$, $D^0(c\bar u)$, or vector mesons $D^{*+}$ at large
values of Feynman variable $x_F$. This may happen either through the coalescence of the valence
and heavy quarks of the same rapidity, or through hadronization of the produced heavy quarks.
In the present paper we do not consider the process of the coalescence (and the corresponding leading
particle effect) and assume that quarks of $c\bar c$-pair fragment into hadrons independently
of each other.

The $x$-dependence for the PDF predicted by the BHPS model is
\begin{equation}
f_c(x,\mu_0) = f_{\bar c}(x,\mu_0) =
 A x^2 \left[ 6x (1+x) \ln x + (1-x)(1+10 x + x^2) \right]. \nonumber
\end{equation}
Here, $\mu_0$ is the starting value for the factorization scale, $\mu_0=m_c$. The
normalization constant $A$ is a variable parameter of the model. The magnitude of IC can be
characterized by the momentum fraction carried by $c + \bar c$,
\begin{eqnarray}
\langle x \rangle_{c+\bar c} = \int\limits_0^1 x \left[ f_c(x,\mu) + f_{\bar c}(x,\mu) \right] dx.
\end{eqnarray}
The global QCD analysis \cite{Pumplin:2007wg} rules out the possibility of an IC component with the momentum fraction
$\langle x \rangle_{c+\bar c}$ much larger $0.02$ (for $\mu=\mu_0=m_c$).

From a point of view of QCD, it is natural to expect that a contribution of
heavy quark component in hadron's wave function cannot be large, due to a suppression by the
power of heavy quark mass squared. However, it is possible that this suppression is compensated
by sufficiently strong nonperturbative gluon fields in the nucleon (see, e.g., \cite{Blotz:1997me}).
It had been shown in \cite{Franz:2000ee} using the technique of heavy quark mass expansion,
that the intrinsic charm contribution to the nucleon momentum, $\langle x \rangle_{c+\bar c}$,
at the scale $\mu \approx m_c$, is proportional to the (unknown) factor $\Lambda$ (which is
determined by the matrix elements of gluon operators) whose value can be obtained using
nonperturbative methods in QCD (e.g., a theory of instanton vacuum). If $\Lambda$ is of
order of typical strong interaction scale, $\Lambda \sim 1$ GeV, then, according to
\cite{Franz:2000ee}, the prediction is
\begin{equation}
\langle x \rangle_{c+\bar c} = {\rm few} \times 10^{-3},
\end{equation}
that is quite similar with the suggestion of \cite{Brodsky:1980pb}.

We consider also, as a second example, the phenomenological model \cite{Pumplin:2007wg}, in which the shape of
the charm PDF is sea-like, i.e., similar to that of the light flavor sea quarks (except for
normalization). Namely, for this case \cite{Pumplin:2007wg}
\begin{eqnarray}
f_{c}(x, \mu_0) = f_{\bar c}(x, \mu_0) \sim f_{\bar d}(x,m_c) + f_{\bar u}(x,m_c).
\end{eqnarray}

It is very important, for high energy cosmic ray physics, that the heavy constituents of an intrinsic
heavy quark state, such as $|uudc\bar c\rangle$, usually carry the largest fraction of the
nucleon's momentum. This can easily be shown in the context of light-cone models, where contributions
of high-mass Fock states (to probability distributions, for example) are suppressed only
by energy denominator factor \cite{Pumplin:2005yf}
\begin{eqnarray}
S = \left[ \sum \limits_{j=1}^{N} \frac{p_{j\perp}^2 + m_j^2}{x_j} - m_p^2 \right] ^ {-2} .
\end{eqnarray}
Here, $j$ is a constituent number, $x_j = p_j^{(+)} /  p_p^{(+)}$ are the light-cone momentum
fractions, $p^{(+)} \equiv (p^{(0)} + p^{(3)})/\sqrt{2} $,
$p^{(0)}$ is a particle's energy and $p^{(3)}$ is its longitudinal momentum (i.e., the momentum
in a direction of the parent nucleon). 
Evidently, if $m_j$ is large, the value
of $x_j$ should also be large, for far off-shell configurations not to be too suppressed.

In two recent works \cite{Goncalves:2008sw, Kniehl:2009ar} the problem of a search of intrinsic
charm was again considered
in connection with experiments at RHIC, LHC and Tevatron. These authors argue that the
hypothesis of intrinsic charm can be checked in these experiments (studying the resulting
transverse momentum spectra of charmed mesons). In the present paper we calculate inclusive
distributions of charmed mesons ($x_F$-spectra) produced in $p-A_{\rm air}$ collisions
at high energies, using the mechanism proposed in \cite{Goncalves:2008sw}
(based on the CGC formalism).

\section{Hadron production in the forward region and CGC}

We are interested in heavy meson production in $pA$-collisions at very high energies and,
what is most essential, at the projectile fragmentation region, i.e., at forward
rapidities. In the forward rapidity region the Bjorken $x$ of the target nucleus has its
lowest possible value for a given $\sqrt{s}$ whereas the Bjorken $x$ of the projectile is
large (close to unity). Rescatterings of the projectile's quarks in a target are quite
substantial. It had been shown in \cite{Dumitru:2002qt, Gelis:2002ki} that a particle production in $p$-nucleus
collisions at high energies and forward rapidities can be expressed through the cross section
of a multiple scattering of a quark from the target nucleus which is treated as a Color
Glass Condensate. In this approach the projectile proton is considered as a collection
of quarks and gluons (according to the parton model), and the standard DGLAP evolution
and collinear factorization are employed. The resulting formula \cite{Dumitru:2005gt} for a single-inclusive
charm production in $pA$ collision is
\begin{eqnarray}
\fl
\frac{d N}{dyd^2p_{\perp}}(pA\to DX) = \frac{1}{(2\pi)^2} \int\limits_{x_F}^{1} dx_p \frac{x_p}{x_F} \times
 \;\;\;\;\;\;\;\;\;  \nonumber \\
\times f_{c/p}(x_p,Q^2) N_F\big(\frac{x_p}{x_F}p_\perp , x_A\big) D_{D/c}
\big(\frac{x_F}{x_p},Q^2 \big).
\label{dNdd}
\end{eqnarray}
Here, $p_\perp$, $y$, $x_F$ are the transverse momentum, rapidity and Feynman $x$ of the produced $D$
meson, respectively. The variable $x_p$ is a momentum fraction of the charm quark of the projectile,
$x_A$ is a momentum fraction of the target parton. The connection between $x_A$ and $x_p$ is
\begin{eqnarray}
x_A = \frac{x_p(p_\perp^2+m_D^2)}{x_F^2 s} = x_p e^{-2 y},
\label{xA}
\end{eqnarray}
and the rapidity is connected with $x_F$, $p_\perp$ by the formula
\begin{eqnarray}
x_F = \frac{\sqrt{p_\perp^2+m_D^2}}{\sqrt{s}} e^y.
\label{xF}
\end{eqnarray}
One can see from (\ref{xA}) and (\ref{xF}) that at large rapidities the $x_A$-values are small while $x_p$ is large.

The functions $f_{c/p}(x_p,Q^2)$ and $D_{D/c}(z,Q^2)$ are the proton charm PDF and
the charm fragmentation function into $D$ mesons, respectively. The factorization scale $Q^2$ is
taken to be equal to $p_\perp^2$. At last, the function $N_F(q_\perp , x_A)$ is equal to 
an imaginary part of the
quark-antiquark dipole-nucleus forward
scattering amplitude, $q_\perp \equiv \frac{x_p}{x_F}p_\perp$. This amplitude 
can be modelled by the following expression \cite{Kharzeev:2004yx}:
\begin{equation} \nonumber
N_F(q_\perp , x_A) \equiv \int d^2 r_\perp e^{i \vec q _\perp \vec r _\perp}  \left[1 -
\exp \left(-\frac{1}{4} \left(r_\perp^2 Q_s^2(x_A) 
\right)^{\gamma(q_\perp,x_A)} \right)\right].
\label{NF}
\end{equation}
Here, $r_\perp$ is the transverse size of the dipole, $Q_s(x)$ is the saturation scale
(see, e.g., \cite{Iancu:2003xm} for a definition of $Q_s$), parameterized by the simple formula 
\begin{eqnarray}
\frac{Q_s^2(x)}{1 \; {\rm GeV}^2} = \frac{N_c^2-1}{2 N_c^2} \left( \frac{x_0}{x} \right)^{\lambda} A^{1/3},
\label{Qs1GeV}
\end{eqnarray}
$x_0 \cong 3\times 10^{-4}$, $\lambda\cong 0.3$, $N_c$ is a number of colours ($N_c=3$).
For the ``anomalous dimension'' factor $\gamma(q_\perp,x_A)$
(which enters Eq. (\ref{NF}) due to quantum evolution in the Color Glass Condensate),
we use the parametrization from \cite{Boer:2007ug}:
\begin{eqnarray}
\gamma(w) = \gamma_1 + (1-\gamma_1) \frac{w^a-1}{(w^a-1)+b},
\end{eqnarray}
where\
\begin{eqnarray}
w=\frac{q_\perp}{Q_s(x_A)},
\end{eqnarray}
and $a=2.82$, $b=168$, $\gamma_1=0.628$. These parameters were fitted to the data of the RHIC
experiment \cite{Arsene:2004ux} where deuteron-gold collisions at forward rapidity (i.e., at 
the fragmentation region of the projectile) were studied. One should note that just this region
of rapidities is essential for calculations in cosmic ray physics.

The formula (\ref{dNdd}) gives the spectrum of produced particles normalized on their average
multiplicity in one collision. Correspondingly, on vertical axes of Figs. \ref{fig1} - \ref{fig3}
we present numbers of produced particles in the $pA$-collision, in unit interval of $x_F$.

For a calculation of the inclusive cross section one must multiply $dN$ in (\ref{dNdd})
on a cross section of the collision, i.e., on a transverse size of the target 
nucleus. To determine an approximate value of the proportionality coefficient,
we note, firstly, that the cross section formula corresponding to Eq. (\ref{dNdd}) is given by
\cite{Dumitru:2005gt, Gelis:2002nn}
\begin{eqnarray}
\fl
\frac{d \sigma (pA\to DX)}{dyd^2p_{\perp}} = \frac{1}{(2\pi)^2} \int\limits_{x_F}^{1} dx_p \frac{x_p}{x_F} f_{c/p}(x_p,Q^2) \times \nonumber \\
\;\;\;\;\;\;\;\;\; \times \int d^2 r_\perp e^{i\vec q_\perp \vec r_\perp} \frac{\sigma_{\rm dip}(r_\perp,x_A)}{2}
\cdot D_{D/c} \big(\frac{x_F}{x_p},Q^2 \big),
\label{dSigmadd}
\end{eqnarray}
where $\sigma_{\rm dip}$ is the dipole-nucleus cross section which, by optical theorem, is expressed
through an imaginary part of the dipole-target forward scattering amplitude (in $r_\perp$-representation):
\begin{eqnarray}
\sigma_{\rm dip}(r_\perp,x) \equiv 2 N_F(r_\perp,x).
\end{eqnarray}
The Fourier-Bessel transformation relates $N_F(r_\perp,x)$ with the amplitude in $q_\perp$-representation,
introduced above, in Eq. (\ref{NF}):
\begin{eqnarray}
\fl
\int d^2 r_\perp e^{i\vec q_\perp \vec r_\perp} N_F(r_\perp,x) = 
 \nonumber \\
\;\;\;\;\;\;\;\;\; = 2 \pi \int r_\perp d r_\perp
J_0(q_\perp r_\perp) N_F(r_\perp,x) \equiv N_F(q_\perp, x).
\label{neweq1}
\end{eqnarray}
For the estimate of the proportionality coefficient,
we use, for simplicity, the quasiclassical approximation, in which anomalous dimension
factor, $\gamma(q_\perp, x)$, is equal to $1$. 
The approximate expression for the dipole cross section can be written as \cite{Bartels:2002cj}
\begin{eqnarray}
\sigma_{\rm dip}(r_\perp,x) = \sigma_0 \left[ 1 -
\exp\left(- \frac{\pi^2 r_\perp^2 \alpha_s(\mu^2)
 x G_A(x,\mu^2)}{3 \sigma_0} \right) \right],
\label{sigma}
\end{eqnarray}
where $\mu^2=\mu_0^2+C/r_\perp^2$, $G_A(x,\mu^2)=AG(x,\mu^2)$ is the gluon density in the nucleus,
$\mu_0^2$, $C$ and $\sigma_0$ are parameters. The expression (\ref{sigma}) agrees, in the
limit of small $r_\perp$, with the leading twist perturbative QCD result. Using, in the
region $\mu^2\sim \mu_0^2$, the definition of the saturation scale $Q_s$ \cite{Iancu:2003xm},
\begin{eqnarray}
Q_s^2(x) = 2 \pi^2 \frac{1}{N_c} \alpha_s(\mu_0^2) \frac{1}{\pi R_A^2} x G_A(x,\mu_0^2),
\label{eqQs2}
\end{eqnarray}
the exponent in (\ref{sigma}) can be rewritten in the form of (\ref{NF}) (with $\gamma\to 1$),
if $\sigma_0=2\pi R_A^2$. 
One can easily check that the Eq. (\ref{eqQs2}) is not in contradiction with the parametrization
of Eq. (\ref{Qs1GeV}) because one has, approximately, 
\begin{eqnarray}
x G_A(x) \sim A x^{-\lambda}, \;\; R_A^2 \sim A^{2/3}.
\end{eqnarray}

Finally, it follows from these considerations that the coefficient, connecting
$d\sigma$ and $dN$, is approximately equal to $\frac{\sigma_0}{2} = \pi R_A^2$.
It is known \cite{Bartels:2002cj} that, for a dipole-nucleon cross section, the value of $\sigma_0$
in Eq. (\ref{sigma}) is equal to $\sim 20\;$mb. This value must be scaled according to
the transverse area of the target nucleus. For the nucleus one has $R_A \approx 1.3 A^{1/3}\;$fm,
and for the nucleon one has $R_N \approx 0.8\;$fm. 

For the numerical calculations in the present paper we used the value of $A$ equal to $14$ having in
mind that this value is close to the average atomic number of the nuclei in atmosphere.
For $A=14$ the proportionality coefficient is 
\begin{eqnarray}
\frac{\sigma_0}{2} = \pi R_A^2 \approx 300 {\rm mb}.
\end{eqnarray}

\section{Results and discussions}

For the concrete calculations we used the charm PDFs presented in \cite{Pumplin:2007wg}.
Authors of \cite{Pumplin:2007wg}
used for the presentation of PDFs at different values of the factorization scale the results of CTEQ
group \cite{Nadolsky:2008zw, Pumplin:2007wg}. For BHPS model, it was
assumed, in the present paper, that the charm content of the proton is on the
maximal level, $\langle x \rangle _ {c+\bar c} = 0.02$. The same is for the sea-like IC model:
$\langle x \rangle _ {c+\bar c} = 0.024$.

\begin{figure}
\centering
\includegraphics[width=0.6 \columnwidth ]{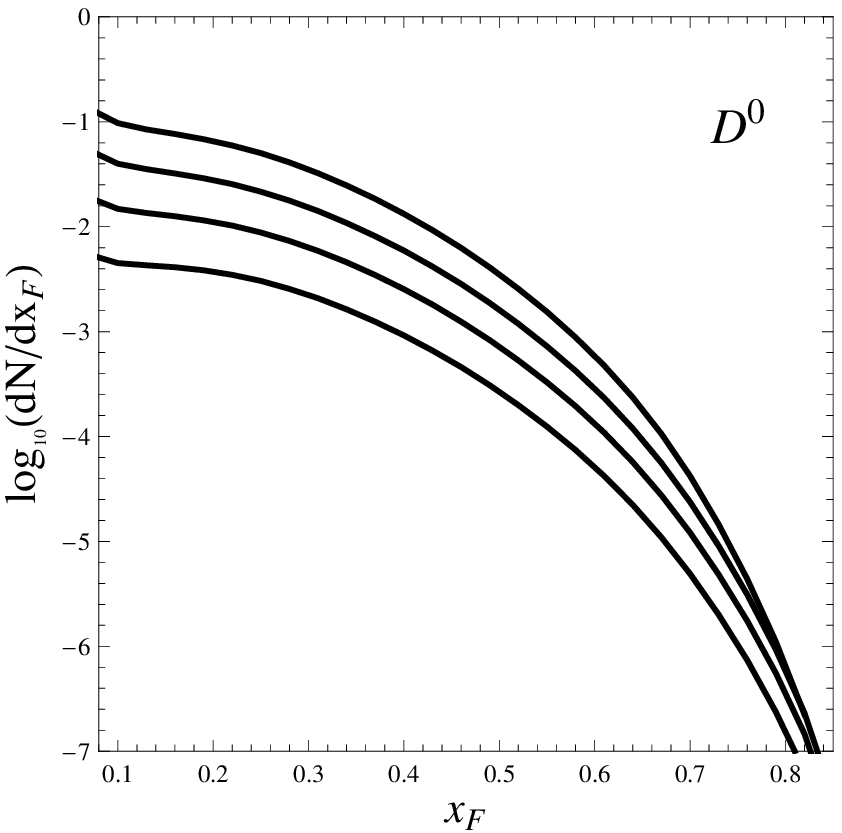}
\caption{The $x_F$-spectra for $D^0$ mesons, $A=14$, for different energies.
$\sqrt{s}=10^2$, $10^3$, $10^4$, $1.7\times 10^5$ GeV (from bottom to top).  }
\label{fig1}
\end{figure}

\begin{figure}
\centering
\includegraphics[width=0.6 \columnwidth ]{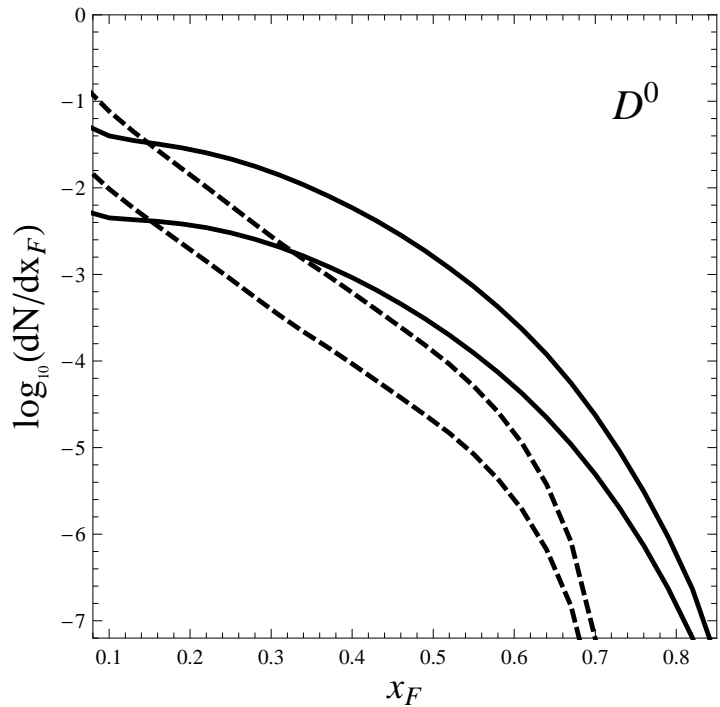}
\caption{The comparison of $x_F$-spectra for BHPS (solid lines) and sea-like (dashed lines)
models, for two energies, $\sqrt{s}=10^2$ and $10^4$ GeV ($A=14$).}
\label{fig2}
\end{figure}

\begin{figure}
\centering
\includegraphics[width=0.6 \columnwidth ]{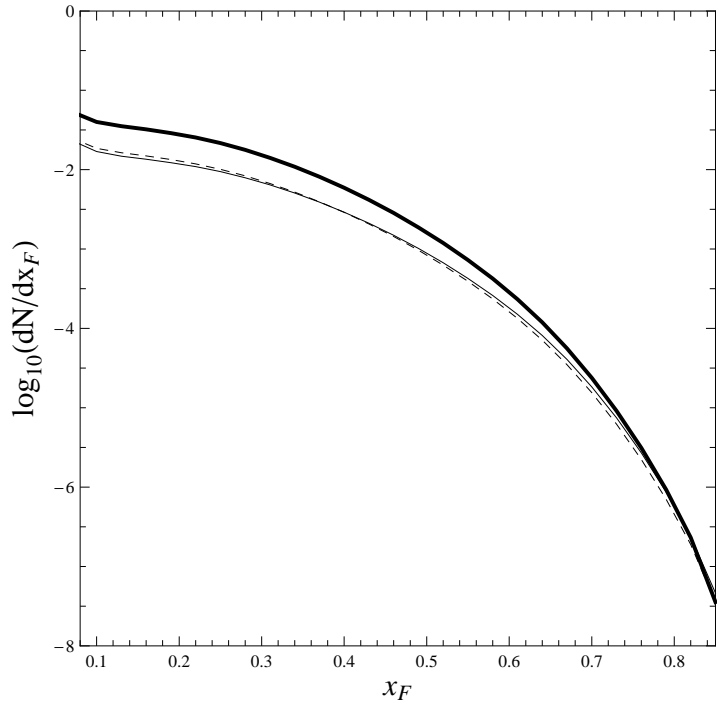}
\caption{The comparison of $x_F$-spectra for different $D$-mesons, for $\sqrt{s}=10^4$ GeV and $A=14$.
Thick solid curve is for $D^0$, dashed curve - for $D^+$, thin solid curve is for $D^{*+}$.}
\label{fig3}
\end{figure}

\begin{figure}
\centering
\includegraphics[width=0.6 \columnwidth ]{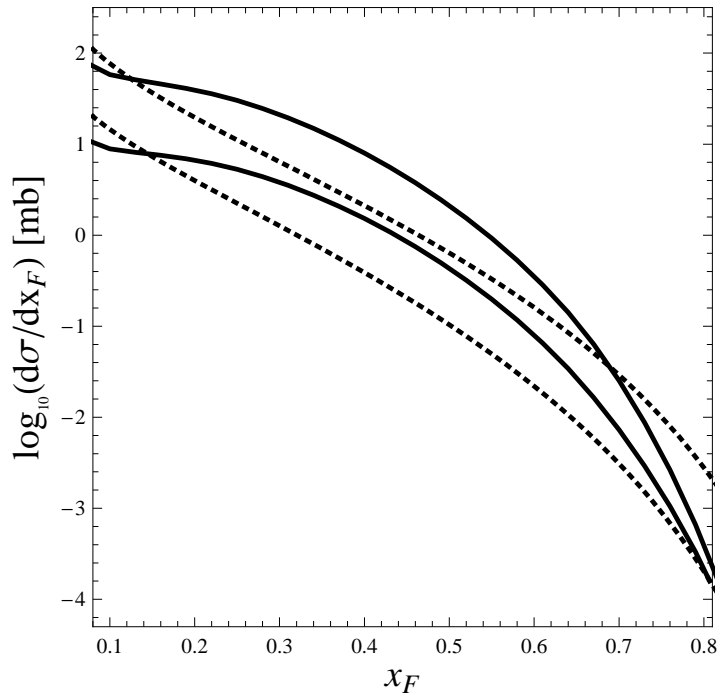}
\caption{The comparison of the present results (solid lines) with the results of
\cite{Goncalves:2006ch} (dashed lines), for $E_p=10^9$ GeV (upper curves) and $3.5 \times 10^4$ GeV
(lower curves) ($A=14$). }
\label{fig4}
\end{figure}

For applications in cosmic ray physics (in particular, for calculations of atmospheric spectra of
leptons (see, e.g., \cite{Bugaev:1989we, Bugaev:1998bi})) we need the dependence of $dN/dx_F$ on $x_F$, so, we
integrate (\ref{dNdd}) on $p_\perp$ and change the variables, using the relation
\begin{eqnarray}
\frac{dx_F}{dy} = x_F.
\end{eqnarray}
The results of the calculations are shown in figures \ref{fig1} - \ref{fig4}.

The $z$ distributions of the $c$-quark fragmentation functions $D(z,Q^2)$ at their starting scales were
assumed to obey the Bowler parametrization \cite{Bowler:1981sb},
\begin{eqnarray}
D(z,\mu_0) = N_c z^{-(1+\gamma_c^2)} (1-z)^{a_c} e^{-\gamma_c^2/z},
\label{Dzmu}
\end{eqnarray}
with three parameters $N_c$, $a_c$, $\gamma_c$. In the present paper we did not calculate the DGLAP
evolution of fragmentation functions, and use (\ref{Dzmu}) for all values of $\mu$.

The concrete values of the parameters $N_c$, $a_c$, $\gamma_c$ were taken from the work
\cite{Kneesch:2007ey}, in which the nonperturbative fragmentation functions for $D$-mesons
had been obtained by fitting experimental data on single-hadron inclusive production in
$e^+e^-$-annihilation obtained by Belle, CLEO, ALEPH and OPAL collaborations.

In figure \ref{fig1} the inclusive spectra of $D^0$-mesons for several values of cms-energy
are shown (for BHPS model). It is seen that the steepness of the spectra strongly varies
with $x_F$: the relatively flat region ends at $x_F\sim 0.5 \div 0.6$. Figure \ref{fig2}, where
the comparison of the BHPS and sea-like models results is given, shows that the form of
$x_F$-spectrum is strongly model-dependent. Figure \ref{fig3} shows that the $x_F$-spectra for
different charmed mesons are quite similar: the slight difference is connected, evidently,
with the corresponding difference in PDFs of light sea antiquarks in the nucleon.

Inclusive cross sections of charmed mesons had been calculated in many works. We compare our
results with the corresponding results of the most recent works \cite{Goncalves:2006ch, Enberg:2008te}.
Authors of \cite{Goncalves:2006ch, Enberg:2008te} calculated the cross sections
of inclusive $c\bar c$-pair production using the
colour dipole-model approach \cite{Nikolaev:1995ty, Raufeisen:2002ka, Kopeliovich:2002yv}
(and extrinsic charm). The results of \cite{Goncalves:2006ch} and \cite{Enberg:2008te} are
quite similar; we compare our curves for $E_p=10^9$ GeV and
$E_p=3.5\times 10^4$ GeV (for BHPS model) with the corresponding
curves of \cite{Goncalves:2006ch} (see figure \ref{fig4}). For this comparison we multiply our
curves for $dN/dx_F$ on $2\cdot \pi R_A^2 \approx 600$ mb.
The coefficient $2$ takes into account, approximately, the contribution of other $D$-mesons
($D^+$ and $D^{*+}$). Figure \ref{fig4} shows that the contribution of the intrinsic charm component
can be relatively large and, in particular, sufficient for applications in cosmic-ray experiments.

It follows from figure \ref{fig4} that in the tail region, $x_F > 0.7$, our curves are even steeper
than those of \cite{Goncalves:2006ch}, in spite of their "intrinsic" origin. One must bear in
mind, however, that in the present paper inclusive spectra of charmed hadrons rather than $c\bar c$
pairs are calculated. Naturally, a taking into account of the fragmentation of $c$ quarks into
hadrons leads to additional steepening of $x_F$-spectra.

\ack
The work was supported by Federal Agency for Science and Innovation under state
contract 02.740.11.5092.

\newpage

\end{document}